%% file: ANA-SUSY-2017-04-PAPER.tex
\pdfoutput=1
\newcommand*{\ATLASLATEXPATH}{}
\documentclass[cernpreprint, texlive=2016, UKenglish]{\ATLASLATEXPATH atlasdoc}
 
\usepackage{\ATLASLATEXPATH atlaspackage}
 
\usepackage{multirow}
\usepackage{rotating}
 
\usepackage[backref=false]{\ATLASLATEXPATH atlasbiblatex}
 
\usepackage[BSM=true]{\ATLASLATEXPATH atlasphysics}
 
\graphicspath{{logos/}{figures/}}
 
\usepackage{ANA-SUSY-2017-04-PAPER-defs}

\AtlasTitle{Search for displaced vertices of oppositely charged leptons from decays of long-lived particles in $pp$ collisions at $\sqrt{s} = 13$~TeV with the ATLAS detector}
 
\AtlasAbstract{
A search for long-lived particles decaying into an oppositely charged lepton pair, $\mu\mu$, $ee$, or $e\mu$, is presented using 32.8~$\mathrm{fb^{-1}}$ of $pp$ collision data collected at $\sqrt{s}=13$~TeV by the ATLAS detector at the LHC.
Candidate leptons are required to form a vertex, within the inner tracking volume of ATLAS, displaced from the primary $pp$ interaction region.
No lepton pairs with an invariant mass greater than 12~GeV are observed, consistent with the background expectations derived from data.
The detection efficiencies for generic resonances with lifetimes ($c\tau$) of 100--1000~mm decaying into a dilepton pair with masses between 0.1--1.0~TeV are presented as a function of $\pt$ and decay radius of the resonances to allow the extraction of upper limits on the cross sections for theoretical models.
The result is also interpreted in a supersymmetric model in which the lightest neutralino, produced via squark--antisquark production, decays into $\ell^{+}\ell^{'-}\nu$ ($\ell, \ell^{'} = e$, $\mu$) with a finite lifetime due to the presence of R-parity violating couplings.
Cross-section limits are presented for specific squark and neutralino masses. For a 700~GeV squark, neutralinos with masses of 50--500~GeV and mean proper lifetimes corresponding to $c\tau$ values between 1~mm to 6~m are excluded. For a 1.6~TeV squark, $c\tau$ values between 3~mm to 1~m are excluded for 1.3~TeV neutralinos.
}
 
\author{The ATLAS Collaboration}

\PreprintIdNumber{CERN-EP-2019-139}
 
\AtlasJournalRef{Phys. Lett. B 801 (2020) 135114}
\AtlasDOI{10.1016/j.physletb.2019.135114}

\hypersetup{pdftitle={ATLAS document},pdfauthor={The ATLAS Collaboration}}
 
\begin{document}
 
\maketitle
 
\section{Introduction}
\label{sec:intro}
 
% The next lines are included from the .//introduction.tex input file
Many extensions to the Standard Model (SM) predict the production of weakly-coupled, long-lived particles (LLPs).
In particular, several models, including supersymmetry (SUSY)~\cite{Golfand:1971iw,Volkov:1973ix,Wess:1974tw,Wess:1974jb,Ferrara:1974pu,Salam:1974ig} with R-parity violation (RPV)~\cite{Dreiner:1997uz,Barbier:2004ez} or with gauge-mediated supersymmetry breaking (GMSB)~\cite{Nappi:1982hm,AlvarezGaume:1981wy,Dine:1981gu}, Hidden Valley models~\cite{Han:2007ae}, dark-photon models~\cite{Holdom:1985ag} or models with long-lived right-handed neutrinos~\cite{Basso:2008iv}, predict the existence of LLPs that can decay into a pair of leptons.
If the LLP has a lifetime of picoseconds to nanoseconds then its decay may be observed as a displaced vertex in the inner tracking volume of the ATLAS detector at the LHC.
 
This letter presents a search for displaced dilepton vertices originating from decays of LLPs into an oppositely charged $\mu\mu$, $ee$, or $e\mu$ pair, with an invariant mass of more than 12~GeV.
The analysis uses data from proton--proton ($pp$) collisions recorded by the ATLAS experiment in 2016 at a centre-of-mass energy of 13~TeV.
Two signal models are used to study the sensitivity of the ATLAS detector to such LLPs.
The first is a simplified RPV SUSY model in which a squark--antisquark pair is produced, each decaying into a long-lived neutralino which results in a pair of charged leptons and a neutrino.
The second is a toy model where a LLP, denoted by $Z'$, is produced in $q\bar{q}$ annihilations and decays into a pair of charged leptons.
These models were selected to study how the kinematic properties of a three- or two-body decay affect the signal efficiencies.
The search is intended to be as model-independent as possible and is not optimized for these particular signal models.
 
Previously, the ATLAS Collaboration searched for displaced dilepton vertices in the inner tracking volume of the ATLAS detector at $\sqrt{s}=8$~TeV~\cite{SUSY-2014-02} and for oppositely charged dimuons using only muon-spectrometer tracks at $\sqrt{s}=13$~TeV~\cite{EXOT-2017-03}.
ATLAS also searched for LLPs with mass of less than 2~GeV at $\sqrt{s}=8$~TeV by considering pairs of highly collimated leptons~\cite{EXOT-2013-22}.
The CMS Collaboration searched for displaced dilepton vertices in the inner tracking volume of the CMS detector at $\sqrt{s}=8$~TeV~\cite{CMS-EXO-12-037} and for electrons and muons with large impact parameters at $\sqrt{s}=8$~TeV~\cite{CMS-B2G-12-024}.
% End of text imported from the .//introduction.tex input file

\section{ATLAS detector}
\label{sec:detector}
 
% The next lines are included from the .//detector.tex input file
\newcommand{\AtlasCoordFootnote}{
ATLAS uses a right-handed coordinate system with its origin at the nominal interaction point (IP) in the centre of the detector and the $z$-axis along the beam pipe.
The $x$-axis points from the IP to the centre of the LHC ring, and the $y$-axis points upwards.
Cylindrical coordinates $(r_{xy},\phi)$ are used in the transverse plane, $\phi$ being the azimuthal angle around the $z$-axis.
The pseudorapidity is defined in terms of the polar angle $\theta$ as $\eta = -\ln \tan(\theta/2)$.
Angular distance is measured in units of $\Delta R \equiv \sqrt{(\Delta\eta)^{2} + (\Delta\phi)^{2}}$.}
 
The ATLAS experiment~\cite{PERF-2007-01,ATLAS-TDR-19,Abbott:2018ikt} at the LHC is a multipurpose particle detector with a cylindrical geometry and a near $4\pi$ coverage in solid angle.\footnote{\AtlasCoordFootnote}
It consists of the inner detector (ID) tracking system surrounded by a thin superconducting solenoid providing a \SI{2}{\tesla} axial magnetic field, electromagnetic and hadronic calorimeters, and a muon spectrometer (MS).
 
The ID extends from a radius of about 33 to 1100~mm and to $|z|$ of about 3100~mm.
It provides tracking for charged particles within the pseudorapidity region $|\eta| < 2.5$.
At small radii, silicon pixel layers and stereo pairs of silicon microstrip detectors provide high-resolution position measurements.
The pixel system consists of four barrel layers, and three forward discs on either side of the ATLAS detector.
The barrel pixel layers, which are positioned at radii of 33.3, 50.5, 88.5, and 122.5~mm are of particular relevance to this search.
The silicon microstrip tracker (SCT) comprises four double layers in the barrel and nine forward discs on either side.
The radial position of the innermost (outermost) SCT barrel layer is 299~mm (514~mm).
The final component of the ID, the transition-radiation tracker (TRT), is positioned at a larger radius, with coverage up to $|\eta|=2.0$.
 
Lead/liquid-argon (LAr) sampling calorimeters provide electromagnetic energy measurements with high granularity.
A steel/scintillator-tile hadron calorimeter covers the central pseudorapidity range ($|\eta| < 1.7$).
The endcap and forward regions are instrumented with LAr calorimeters for electromagnetic and hadronic energy measurements up to $|\eta| = 4.9$.
 
The MS surrounds the calorimeters and is immersed in a toroidal magnetic field.
It provides tracking for charged particles within the pseudorapidity region $|\eta| < 2.7$ and trigger information up to $|\eta| = 2.4$.
Three layers of muon detectors are arranged in concentric shells at radii between 5~m and 10~m in the barrel region, and in wheels perpendicular to the beam axis at radii between 7.4~m and 21.5~m in the endcap regions where $|\eta| > 1.05$.
 
A two-level trigger system~\cite{TRIG-2016-01} is used to select events.
The first-level trigger is implemented in hardware and uses a subset of the detector information to reduce the accepted rate to at most \SI{100}{\kilo\hertz}.
This is followed by a software-based, high-level trigger that reduces the accepted event rate to about \SI{1}{\kilo\hertz}.
% End of text imported from the .//detector.tex input file

\section{Dataset and simulated events}
\label{sec:dataMC}
 
% The next lines are included from the .//dataMC.tex input file
This analysis uses the $pp$ collision data recorded in 2016, corresponding to an integrated luminosity of 32.8~$\mathrm{fb^{-1}}$ obtained while all parts of the detector were operational.
The uncertainty in the integrated luminosity is 2.2\%~\cite{ATLAS-CONF-2019-021}, obtained using the LUCID-2 detector~\cite{Avoni:2018iuv} for the primary luminosity measurements.
 
The sensitivity of the ATLAS detector to LLPs decaying into a pair of leptons was studied using Monte Carlo (MC) simulations of two different signals.
In the RPV SUSY simplified model, a pair of left- and right-handed squarks of the first two generations was produced via the strong interaction.
These eight squarks were assumed to be mass-degenerate and decay into their SM partner and the lightest supersymmetric particle (LSP), which is a bino-like neutralino ($\ninoone$).
All other SUSY particles were assumed to be decoupled.
The LSP decay was mediated by the following lepton-number-violating superpotential term~\cite{Barbier:2004ez}:
\begin{equation}
\label{LLE}
W_{\text{LLE}} = \frac{1}{2} \lambda_{ijk} L_{i} L_{j} \bar{E}_{k},
\end{equation}
where $\lambda_{ijk}$ are the coupling strengths of the interactions; $i$, $j$, $k$ denote the fermion generation; and $L$ and $\bar{E}$ are $\text{SU}(2)$-doublet and singlet superfields, respectively, which contain the SM leptons and their superpartners.
In the presence of these interactions, the LSP decays via phase space into a lepton and a virtual slepton, whose decay is described by Eq.~\eqref{LLE}, resulting in the following three-body decay:
\begin{equation}
\notag
\ninoone \to \ell_{k}^{\pm} \ell_{i/j}^{\mp} \nu_{j/i}.
\end{equation}
 
This model was interpreted in two different scenarios, where a single dominant $\lambda_{121}$ or $\lambda_{122}$ coupling was assumed.
Decays via a pure $\lambda_{121}$ coupling have branching fractions $\mathcal{B}(\ninoone \to ee\nu) = \mathcal{B}(\ninoone \to e\mu\nu) = 0.5$, while decays via a pure $\lambda_{122}$ coupling have $\mathcal{B}(\ninoone \to \mu\mu\nu) = \mathcal{B}(\ninoone \to e\mu\nu) = 0.5$.
 
The MC samples of the RPV SUSY model were generated with \MGMCatNLOV{2.2.3}~\cite{Alwall:2014hca} interfaced to \PYTHIAV{8.210}~\cite{Sjostrand:2014zea} using the A14 set of underlying event and hadronization parameters (A14 tune)~\cite{ATL-PHYS-PUB-2014-021} and the NNPDF23LO PDF set~\cite{Ball:2012cx}.
The matrix-element calculation was performed at the tree level and included the emission of up to two additional partons.
The merging of matrix elements and parton showers was done with the CKKW-L algorithm~\cite{Lonnblad:2011xx}, with a matching scale set to one quarter of the squark mass.
The cross-sections in the SUSY model were calculated to approximate, next-to-next-to-leading order in the strong coupling constant, including the resummation of soft gluon emission at next-to-next-to-leading-logarithmic accuracy (approximate NNLO+NNLL)~\cite{Beenakker:2016lwe,Beenakker:2014sma,Beenakker:2013mva,Beenakker:2011sf,Beenakker:2009ha,Kulesza:2009kq,Kulesza:2008jb,Beenakker:1996ch}.
The nominal cross-section and its uncertainty were derived using the PDF4LHC15\_mc PDF set, following the recommendations of Ref.~\cite{Butterworth:2015oua}.
The samples span four hypotheses for the squark and the LSP masses and mean proper lifetimes corresponding to $c\tau =$~10--1000~mm.
Instead of producing MC samples for each of the two scenarios separately, combined samples were produced in which the LSP decays into the three final states ($ee\nu$, $e\mu\nu$ and $\mu\mu\nu$) at the same rate.
The events were reweighted to match the chosen $\lambda$ coupling.
 
The events of each MC sample of the RPV SUSY model were generated for a specific mean proper lifetime $\tau_{\text{MC}}$ of the LSP.
To obtain the signal efficiency for a different mean proper lifetime $\tau$, a weight was given to each LSP which decayed at the time $t$:
\begin{equation}
\notag
w_{\text{LSP}} = \frac{\tau_{\text{MC}}}{\tau} \exp \left( \frac{t}{\tau_{\text{MC}}} - \frac{t}{\tau} \right).
\end{equation}
The final prediction of the signal efficiency at a given lifetime and mass was determined by applying this reweighting procedure individually to all MC samples of different lifetimes but the same mass, and calculating the weighted mean for the efficiency.
 
The second signal is a toy model of a $Z'$ boson with mass of 100 to 1000~GeV and mean proper lifetime corresponding to a $c\tau$ value of 100, 250, or 500~mm.
The natural width was based on a relativistic Breit--Wigner distribution, and was varied from 10 to 140~GeV.
The samples were generated with \PYTHIAV{8.212} using the A14 tune and the NNPDF23LO PDF set.
Such a directly produced $Z'$ is expected to be excluded by searches for displaced hadronic jets, since it would decay into $q\bar{q}$ with a high branching fraction.
The purpose of this model is to derive efficiencies for a two-body decay of an LLP that can be applied to other models of similar kinematics.
 
In this analysis, all backgrounds were estimated from data.
A selection of SM MC samples were used to test and validate the background estimation techniques for random crossings of tracks in two-track vertices and to estimate the systematic uncertainties in vertexing and tracking.
These included MC samples of $t\bar{t}$ events generated using \sherpa~2.2~\cite{Gleisberg:2008ta} with the NNPDF30NNLO PDF set~\cite{Ball:2014uwa}, leptonic decays of dibosons ($ZZ$, $WW$, $WZ$) from \sherpa~2.1 with the CT10 PDF set~\cite{Lai:2010vv}, and dijet events generated using \PYTHIAV{8.186}~\cite{Sjostrand:2006za,Sjostrand:2007gs} with the A14 tune and the NNPDF23LO PDF set~\cite{Ball:2012cx}.
These MC samples contain isolated leptons (from gauge boson decays) that have a high transverse momentum $\pt$ and leptons and displaced tracks in $b$-jets, in addition to tracks from pile-up vertices for the study of background and systematic uncertainties.
 
The ATLAS detector simulation~\cite{SOFT-2010-01} is based on \GEANT~4~\cite{Agostinelli:2002hh}.
The generation of the simulated event samples included the effect of multiple $pp$ interactions per bunch crossing, as well as the effect on the detector response due to interactions from bunch crossings before or after the one containing the hard interaction.
Multiple $pp$ collisions per bunch crossing were simulated with the soft QCD processes of \PYTHIAV{8.186} using the A2 tune~\cite{ATLAS:2012uec} and the MSTW2008LO PDF set~\cite{Martin:2009iq}.
% End of text imported from the .//dataMC.tex input file
 
\section{Reconstruction and event selection}
\label{sec:reco_sel}
 
% The next lines are included from the .//recon_selection.tex input file
Measurements in the ID are used to reconstruct tracks and vertices, while information from the calorimeters and the muon spectrometer is used to identify electron and muon candidates.
The analysis imposes selection criteria on the two leptons but not on other collision activity, except for requiring the presence of a primary vertex (PV), a vertex with at least two tracks and the highest scalar sum of the $\pt^2$ of associated tracks..
This ensures that the analysis is sensitive to a wide range of new physics models.
Events must also satisfy the standard data-quality criteria of ATLAS and must have a PV with $z$ location that satisfies $|z_{\text{PV}}| < 200$~mm.
The latter requirement has a negligible impact on signal efficiency.
 
\subsection{Triggers and preselection}
\label{sec:presel}
 
In order to improve the sensitivity to decays of LLPs, dedicated algorithms are used for the reconstruction of tracks and vertices.
As these algorithms are CPU intensive and significantly increase the storage size of a recorded collision event, they are applied only to a subset of the data events identified using a set of preselection criteria.
 
The preselection criteria are based on triggers that do not use ID track information, because the track reconstruction applied at the trigger stage does not reconstruct tracks from particles produced with significant displacement from the $pp$ interaction vertex.
Displaced muon candidates are selected with a muon trigger requiring a MS track that has $\pt > 60$~GeV and $|\eta| < 1.05$.
Photon triggers are used to select displaced electron candidates, triggering on the electromagnetic showers produced by the electrons.
These include a single-photon trigger with photon $\pt > 140$~GeV and a diphoton trigger requiring both photons to have $\pt > 50$~GeV.
The trigger efficiency is 1--90\%, depending on the mass and lifetime of the LLP.
 
In order to be preselected, events that pass the triggers are required to satisfy an additional set of criteria, which is evaluated using the data reconstructed with the standard ATLAS algorithms, including the electron~\cite{Aaboud:2019ynx} and muon reconstruction algorithms~\cite{Aad:2016jkr}.
The preselected events are stored in a raw data format so that the reconstruction optimized for LLP decays can be performed.
 
The preselection criteria for muon and electron candidates are summarized in Table~\ref{table:presel_criteria}.
If the muon is reconstructed using an ID track and there is a good match with an MS track based on the $\chi^{2}$ per degree of freedom, $\chi^{2}/\text{DoF} < 5$, then the transverse impact parameter, calculated relative to the beamline, must satisfy $|d_0|$ > 1.5~mm.
The preselection criteria for electron candidates depend on whether a single or diphoton trigger is passed.
If a photon instead of an electron satisfies the single photon trigger, an additional candidate object is required to further limit the size of the data set.
The reconstructed photons are required to pass the `loose' photon identification criteria~\cite{Aaboud:2018yqu}.
 
\begin{table}[tbp]
\centering
\caption{Requirements on the muon, photon and electron candidates that pass the triggers used in the preselection. $e^{*}$ refers to electrons that are required to pass the `loose' electron identification criteria~\cite{Aaboud:2019ynx}.}
\label{table:presel_criteria}
\begin{tabular}{c|cccc|cccc}
\hline
\hline
Trigger &
Candidate 1 & \pt{}~[\GeV] & \abseta{} & $|d_0|$~[mm] &
Candidate 2 & \pt{}~[\GeV] & \abseta{} & $|d_0|$~[mm] \\
\hline
$\mu$ & $\mu$ & $> 62$ & < 1.07 & see text & & & \\
\hline
$\gamma$ & $\gamma$ & $>150$ & $< 2.5$ & -- & $\gamma$ & $>10$ & $< 2.5$ & -- \\
& & & & & $e$ & $>10$ & $< 2.5$ & $>2.0$ \\
& & & & & $\mu$ & $>10$ & $< 2.5$ & see text \\
$\gamma$ & $e$ & $>150$ & $< 2.5$ & $>2.0$ & & & & \\
\hline
$\gamma\gamma$ & $\gamma$ & $>55$ & $< 2.5$ & -- & $\gamma$ & $>55$ & $< 2.5$ & -- \\
$\gamma\gamma$ & $e$ & $> 55$ & $< 2.5$ & $> 2.0$ & $e$ & $> 55$ & $< 2.5$ & $> 2.0$ \\
$\gamma\gamma$ & $e^{*}$ & $> 55$ & $< 2.5$ & -- & $\gamma$ & $> 55$ & $< 2.5$ & -- \\
$\gamma\gamma$ & $e$ & $> 55$ & $< 2.5$ & $> 2.0$ & $\gamma$ & $> 55$ & $< 2.5$ & -- \\
\hline
\hline
\end{tabular}
\end{table}
 
\subsection{Reconstruction of long-lived particle decays}
\label{sec:reco}
 
The standard tracking (ST) and vertex reconstruction in ATLAS are optimized for the reconstruction of particles that originate near the $pp$ interaction vertex.
The ST requirements on the transverse ($d_{0}$) and longitudinal ($z_{0}$) impact parameters, calculated relative to the beamline and beam spot respectively, limit the reconstruction efficiency for highly displaced decays of LLPs.
In order to improve the efficiency, a second track reconstruction algorithm, large radius tracking (LRT)~\cite{ATL-PHYS-PUB-2017-014}, as well as a dedicated displaced-vertex (DV) reconstruction is used.
 
The track requirements in the ST and LRT are compared in Table~\ref{table:lrt_comparison}.
Of particular note is the relaxation of the requirements on $d_{0}$, $z_{0}$, and the maximum number of shared silicon modules and the minimum number of unshared silicon hits between two tracks.
The LRT reconstructs tracks based on the remaining hits that were not used by the ST, and the resulting tracks from both track reconstruction algorithms are used in the analysis.
 
\begin{table}[tbp]
\centering
\caption{Comparison of track requirements between the standard and large radius tracking.}
\label{table:lrt_comparison}
\begin{tabular}{ l  r  r }
\hline
\hline
& Standard & Large radius \\ [0.5ex]
\hline
Maximum $|d_{0}|$ [mm] & 10 & 300 \\
Maximum $|z_{0}|$ [mm] & 250 & 1500 \\
Maximum $|\eta|$ & 2.7 & 5 \\
Maximum shared silicon modules & 1 & 2 \\
Minimum unshared silicon hits& 6 & 5 \\
Minimum silicon hits & 7 & 7\\
\hline
\hline
\end{tabular}
\end{table}
 
The tracks used in the DV reconstruction must satisfy the following requirements: $\pt > 1.0$~GeV, $2 < |d_0| < 300$~mm, and $|z_0| < 1500$~mm.
In addition, each track must have at least two SCT hits and either at least two pixel hits or at least one TRT hit, with no more than two silicon hits shared with another track.
The selected ID tracks are used to find two-track seed vertices.
The vertex tracks are not allowed to have hits in pixel layers at radii smaller than the seed vertex, and are required to have a hit in the nearest pixel or SCT layer at larger radius.
Multi-track DVs are formed iteratively using the collection of seed vertices.
In this process, DVs that are close to each other are merged and poorly associated tracks are removed from DVs to ensure that each selected track is assigned to only one DV~\cite{SUSY-2016-08}.
 
The performance and uncertainties associated to the LRT and vertexing are studied with reconstructed $\Kshort$, as described in Section~\ref{sec:systematics}.
 
\subsection{Displaced-vertex selection}
\label{sec:vtx_sel}
 
DV candidates are required to satisfy several selection criteria in order to select DVs with good efficiency and low fake rate.
The fit of each DV is required to have $\chi^2 /\text{DoF} < 5$ to reject poorly reconstructed vertices.
A minimum transverse displacement, $d_{xy} = \sqrt{(x_{\text{DV}} - x_{pp})^{2} + (y_{\text{DV}} - y_{pp})^{2}} >$~2~mm, to all vertices from proton--proton collisions ($x_{pp}$, $y_{pp}$) is required in order to suppress background from prompt decays.
DVs are selected only if their transverse radius $r_{xy} = \sqrt{x_{\text{DV}}^{2} + y_{\text{DV}}^{2}}$ is less than 300~mm and $|z_{\text{DV}}| < 300$~mm (`fiducial volume').
The invariant mass of the tracks forming the DV must satisfy $m_{\text{DV}} > 12$~GeV to suppress backgrounds from low-mass SM particles.
Each DV is required to have at least one positively charged track and at least one negatively charged track.
DVs are rejected if they are within the volume directly in front of disabled pixel modules~\cite{SUSY-2016-08}, which reduces the fiducial volume by 2.3\%.
DVs with electrons that originate within tracking layers or support structures~\cite{SUSY-2016-08} are rejected (`material veto') to eliminate any remaining background not vetoed by the invariant mass requirement.
About 42\% of the fiducial volume is discarded by the material veto.
 
Each DV has to be associated with at least two leptons (see below).
Non-leptonic DV tracks are not rejected and have to satisfy only the track requirements applied in the DV reconstruction.
A DV associated with both electrons and muons is classified as an $e\mu$ vertex, while DVs associated with electrons or muons only are classified as $ee$ or $\mu\mu$ vertices, respectively.
One of the leptons of a DV is required to satisfy the trigger and preselection criteria discussed in Section~\ref{sec:presel}, except in the case of the diphoton trigger, where both leptons must satisfy the preselection criteria.
 
\subsection{Lepton identification and selection}
\label{sec:lepton_reconstruction}
 
In this search, the standard ATLAS electron and muon reconstruction and identification algorithms~\cite{Aaboud:2019ynx, Aad:2016jkr} are slightly modified to improve the efficiency at large impact parameters.
In particular, any requirement on the number of pixel hits is removed.
Leptons are reconstructed using ID tracks from both the ST and the LRT.
Reconstruction ambiguities between electrons and muons are removed to ensure that an ID track is associated with only one muon or electron.
In this procedure, any electron sharing an ID track with a muon is removed.
In addition, the object with lower $\pt$ is removed if two electrons or two muons share an ID track.
 
The electron candidates are identified using a modified version of the `loose' operating point of the electron identification~\cite{Aaboud:2019ynx}, where the $d_0$ information is not used in the calculation of the likelihoods.
Electrons are rejected if they are associated with an energy cluster in the calorimeter that does not satisfy standard quality criteria.
Muons are reconstructed~\cite{Aad:2016jkr} by matching MS tracks to ID tracks and by performing a global re-fit of all measurements.
For muons with $0.1 < |\eta| < 1.9$, at least 10\% of the TRT hits originally assigned to the ID track have to be included in the final fit~\cite{Aad:2016jkr}.
Both the electrons and muons are required to have $\pt > 10$~GeV.
In addition, electrons (muons) have to satisfy $|\eta| < 2.47$~($2.5$).
 
A cosmic-ray muon passing through the ID during a collision event can be reconstructed as a pair of muons if one muon track segment is reconstructed in the wrong direction (i.e. opposite to its true direction of motion).
The two reconstructed muons will appear to have opposite electric charges and form a high-mass DV.
These back-to-back muons have $|\Delta\phi| \approx \pi$ and opposite $\eta$'s, $\eta_1 + \eta_2 \approx 0$.
Cosmic-ray muons are therefore effectively rejected with a veto, $\Delta R_{\mathrm{cos}} = \sqrt{(|\Delta\phi| - \pi)^{2} + (\eta_1 + \eta_2)^{2}} < 0.01$.
This veto is applied to any type of lepton pair in case a cosmic-ray muon is reconstructed as an electron.
Only events with a single cosmic-ray muon are considered in the background estimation, as the possibility of having multiple cosmic-rays in an event was found to be negligible.
% End of text imported from the .//recon_selection.tex input file
 
\section{Background estimation}
\label{sec:background}
 
% The next lines are included from the .//background.tex input file
Owing to the DV mass requirement ($m_{\text{DV}} > 12$~GeV) low-mass SM processes are an insignificant source of background.
The displacement requirements ($d_{xy} > 2$~mm and $|d_0| > 2$~mm) further reduce the contributions from other SM processes to negligible levels.
Consequently, the largest background source is cosmic-ray muons and the random crossing of two uncorrelated leptons as discussed below.
 
\begin{figure}[tbp]
\centering
\includegraphics[width = 0.7\textwidth]{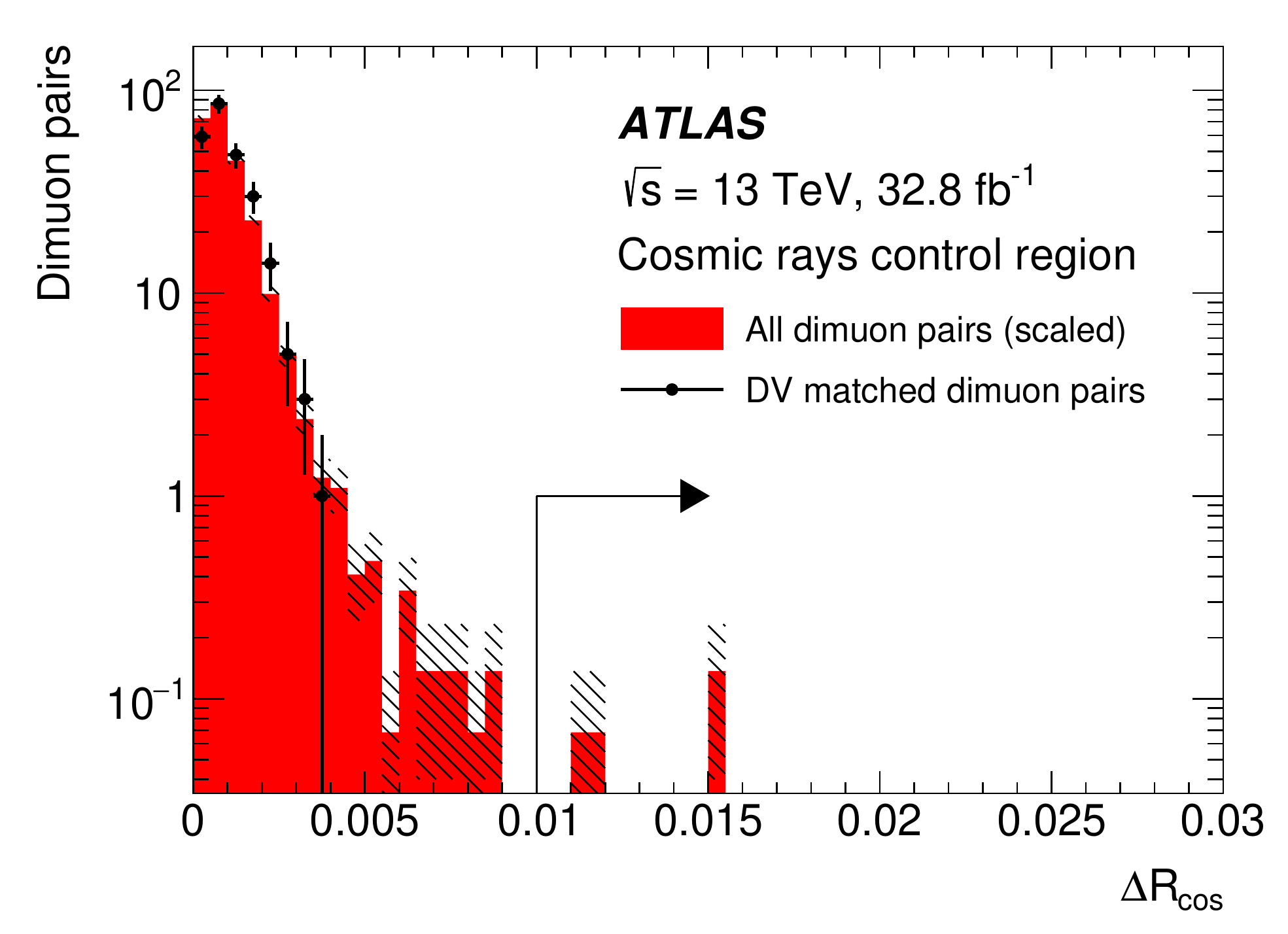} \\
\caption{$\DRcos$ for dimuon events observed in the cosmic rays control region in collision data. The filled histogram includes all dimuon pairs, with the statistical uncertainties represented by the shaded bands. The dots indicate the subset forming a DV candidate. The distribution of all dimuon pairs is scaled to the DV distribution with $\DRcos < 0.004$. The arrow illustrates the signal region requirement of $\DRcos > 0.01$.}
\label{fig:cosmicCR}
\end{figure}

The cosmic-ray background is estimated in a cosmic rays control region consisting of events with back-to-back muons.
There are 3646 dimuons in this control region of which 246 dimuons form a displaced vertex.
Figure~\ref{fig:cosmicCR} shows the $\DRcos$ distribution of the two samples with the former normalized to the latter using the integrals over $\DRcos < 0.004$.
The background estimate is given by the number of dimuons after normalization that satisfy the $\DRcos > 0.01$ requirement of the signal region.
 
The procedure yields a background estimate of $N_{\text{bkg}}^{\text{cos}} = 0.27 \pm 0.14 \textrm{ (stat.)} \pm 0.10 \textrm{ (syst.)}~\text{events}$ in the combined ($ee$, $e\mu$, $\mu\mu$) signal region.
The systematic uncertainty is estimated by alternatively using the first three bins ($\DRcos < 0.0015$) for the normalization and evaluating the difference in the region $0.0015 < \DRcos < 0.004$.
The signal contamination in the cosmic rays control region has been studied using the two signal models and found to be negligible.
 
To estimate the background contribution from randomly crossing tracks and/or leptons, two data-driven methods are used; \textit{event mixing} and \textit{track flipping}.
The \textit{event mixing} method randomly samples leptons from different events to measure how often a pair would form a vertex.
The \textit{track flipping} method performs secondary-vertex reconstruction on events after one track is randomly selected and flipped relative to the beam spot\footnote{In the track flipping, the track parameters are transformed as: $d_0 \to - d_0$, $z_0 \to - z_0$, $\theta \to \pi - \theta$, $\phi \to \phi - \pi$ for $\phi > 0$ and $\phi \to \phi + \pi$ for $\phi \leq 0$.}.
 
The background estimate from the event mixing method is obtained by multiplying the number of track pairs observed in data by the probability that the two tracks will form a displaced vertex by chance.
The sum of the background in the three dilepton samples is estimated to be $2.4 \times 10^{-3}$~events.
The track flipping method estimates a comparable background, $3.9 \times 10^{-3}$~events.
Both methods are tested with a MC sample of $t\bar{t}$ events using tracks without requiring lepton identification and yield about 20\% more random crossing background.
The final estimate uses the event mixing method because it relies on fewer assumptions in the background estimation, and the difference between these two methods, of 64\%, is used as a systematic uncertainty.
This yields a background estimate of $N_{\text{bkg}}^{\text{cross}} = 0.0024 \pm 0.0005 \textrm{ (stat.)} \pm 0.0015 \textrm{ (syst.)}~\text{events}$.
This subdominant background contribution is neglected in final interpretations.
% End of text imported from the .//background.tex input file
 
\section{Signal systematic uncertainties}
\label{sec:systematics}
 
% The next lines are included from the .//systematic_uncertainties.tex input file
Five systematic uncertainties in the signal yields are considered.
The uncertainties in the production cross-section of the SUSY model are 8.7\% for a 700~GeV squark and 17.8\% for a 1.6~TeV~squark.
The systematic uncertainty in the integrated luminosity of 32.8~$\mathrm{fb^{-1}}$ is 2.2\%, as explained in Section~\ref{sec:dataMC}.
 
The simulated events are reweighted to reproduce the distribution of the mean number of interactions per bunch crossing ($\mu$) in the data.
In this procedure, the $\mu$ measured in the data is divided by a factor of $1.09 \pm 0.09$, which improves the agreement between the data and simulation for the observed number of primary vertices and reproduces the fraction of the visible cross-section of inelastic $pp$ collisions as measured in the data~\cite{Aaboud:2016mmw}.
The resulting uncertainty in the signal efficiency is usually less than 10\%, but can be larger in cases with a small number of events in the MC simulations.
 
The trigger efficiencies in the simulations are scaled to those measured in the data using scale factors, which are obtained from a tag-and-probe method applied to $Z$+jets events.
In this method, a pair of leptons is selected, where the tag lepton satisfies tight identification and isolation criteria and the probe lepton satisfies the same identification criteria as applied in the signal region.
No requirements on the impact parameters of the probe and tag leptons are made.\footnote{Studies in signal MC samples have shown that a dependence of the trigger efficiencies on the impact parameters of the leptons can be neglected.}
The requirements on the tag lepton and the invariant mass of the pair ($|m(\ell^{+}\ell^{-}) - m(Z)| < 10$~GeV) ensure that lepton pairs from $Z$ boson decays are selected with a low background, while the probe lepton is used to determine the trigger efficiency.
The scale factors are derived with uncertainties of a few percent by comparing the trigger efficiencies measured in the data with those of a $Z$+jets MC sample.
The average scale factor is 0.98--0.99 for the photon triggers and 0.88 for the muon trigger.
 
The decay $\Kshort\rightarrow\pi^{+}\pi^{-}$ is used to study the systematic uncertainty in the LRT and vertexing, since this sample provides adequate statistics.
The $\Kshort$ sample is selected from the same data set as used in the LLP search with the same triggers.
Tracks originating from a $\Kshort$ decay can be reconstructed by either the ST or the LRT algorithm.
The systematic uncertainties associated with the standard track reconstruction~\cite{ATL-PHYS-PUB-2015-051} and the secondary vertexing~\cite{PERF-2015-06} without the LRT are well understood.
Systematic uncertainties due to the LRT are estimated by examining the differences in the $\Kshort$ yield between data and MC simulation as a function of vertex radius.
 
Events have to satisfy the event selection criteria described above except that the preselection is not explicitly required, as the high-$\pt$ photon or muon triggers would reduce the statistical precision of this study significantly.
Each $\Kshort$ candidate vertex must have exactly two tracks that are oppositely charged and an invariant mass between 350 and 650~MeV.
They have to satisfy the standard vertex selection criteria described in Section~\ref{sec:vtx_sel}.
The difference of the longitudinal impact parameters between the two tracks, each relative to the $\Kshort$ vertex, is required to be less than 2~mm.
 
The numbers of $\Kshort$ vertices found in the data and background MC samples are binned in transverse vertex radius, $r_{xy}$.
The $\Kshort$ yields in each bin are estimated after subtracting the background contributions using sidebands in the invariant mass distribution.
The vertex yields of $\Kshort$ obtained with LRT are compared between the data and the background MC sample, where the data are normalized by adjusting the ST $\Kshort$ yield in data to that of the MC simulation.
The largest difference in the $r_{xy}$ distribution between data and MC simulation is chosen as the systematic uncertainty, and the statistical uncertainty in the difference is then included to yield a total systematic uncertainty of 6\%.
 
In addition to the systematic uncertainty of 6\% from LRT and vertexing, there is a systematic uncertainty of 2\% from ST~\cite{ATL-PHYS-PUB-2015-051} and 1\% from varying the parameters of the secondary vertexing~\cite{PERF-2015-06}.
The total uncertainty from tracking and vertexing is 10\%, obtained by adding all the contributions linearly.
 
The uncertainty on the electron or muon identification efficiency can be neglected in comparison to the uncertainty in the LRT and vertexing.
% End of text imported from the .//systematic_uncertainties.tex input file
 
\section{Results and interpretations}
\label{sec:result}
 
% The next lines are included from the .//results.tex input file
No events are observed satisfying the selection criteria of the signal region, consistent with the estimated background of $0.27 \pm 0.17$ events.
 
\begin{figure}[tbp]
\centering
\subfloat{\includegraphics[width=0.5\textwidth]{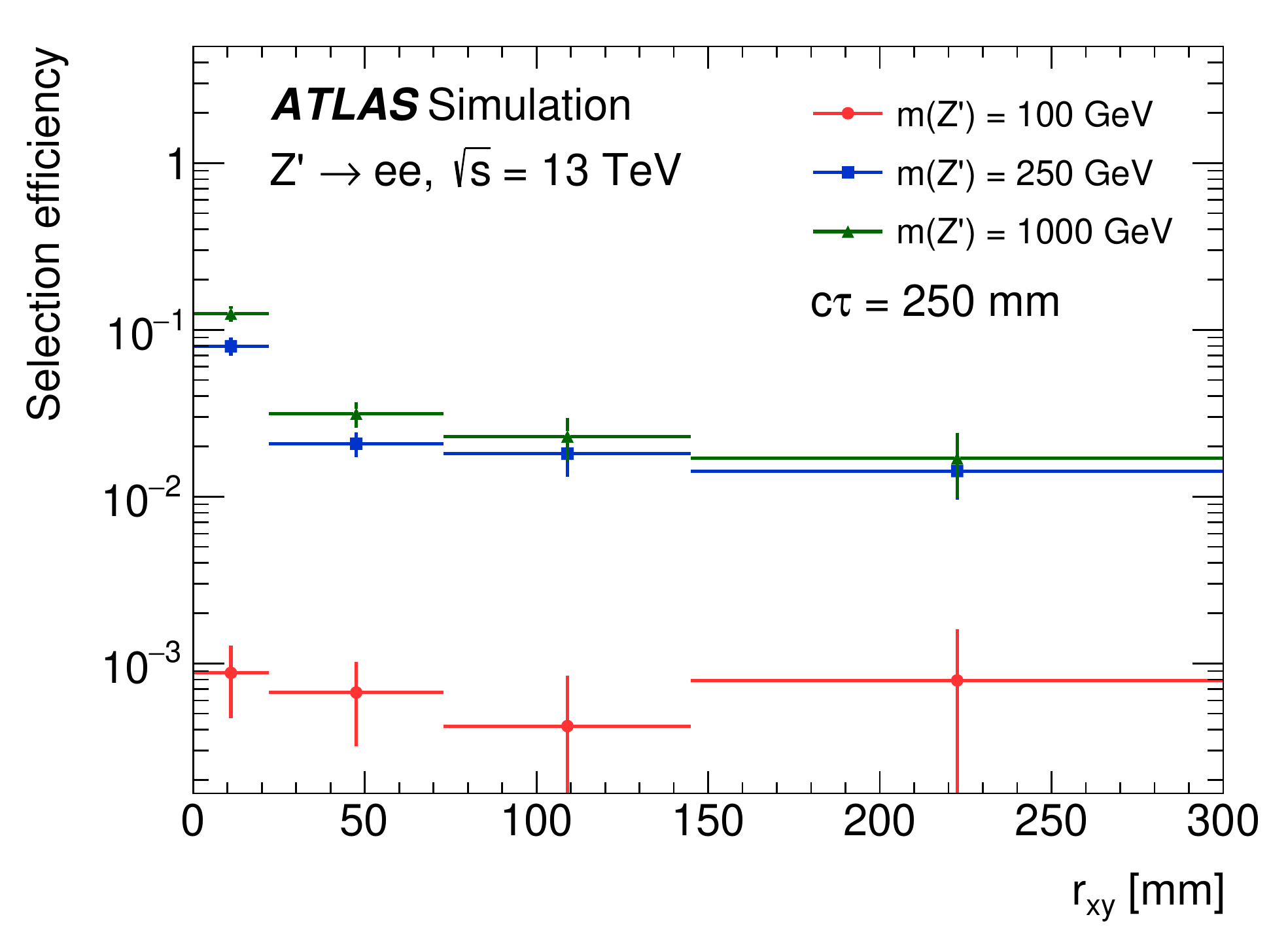}}
\subfloat{\includegraphics[width=0.5\textwidth]{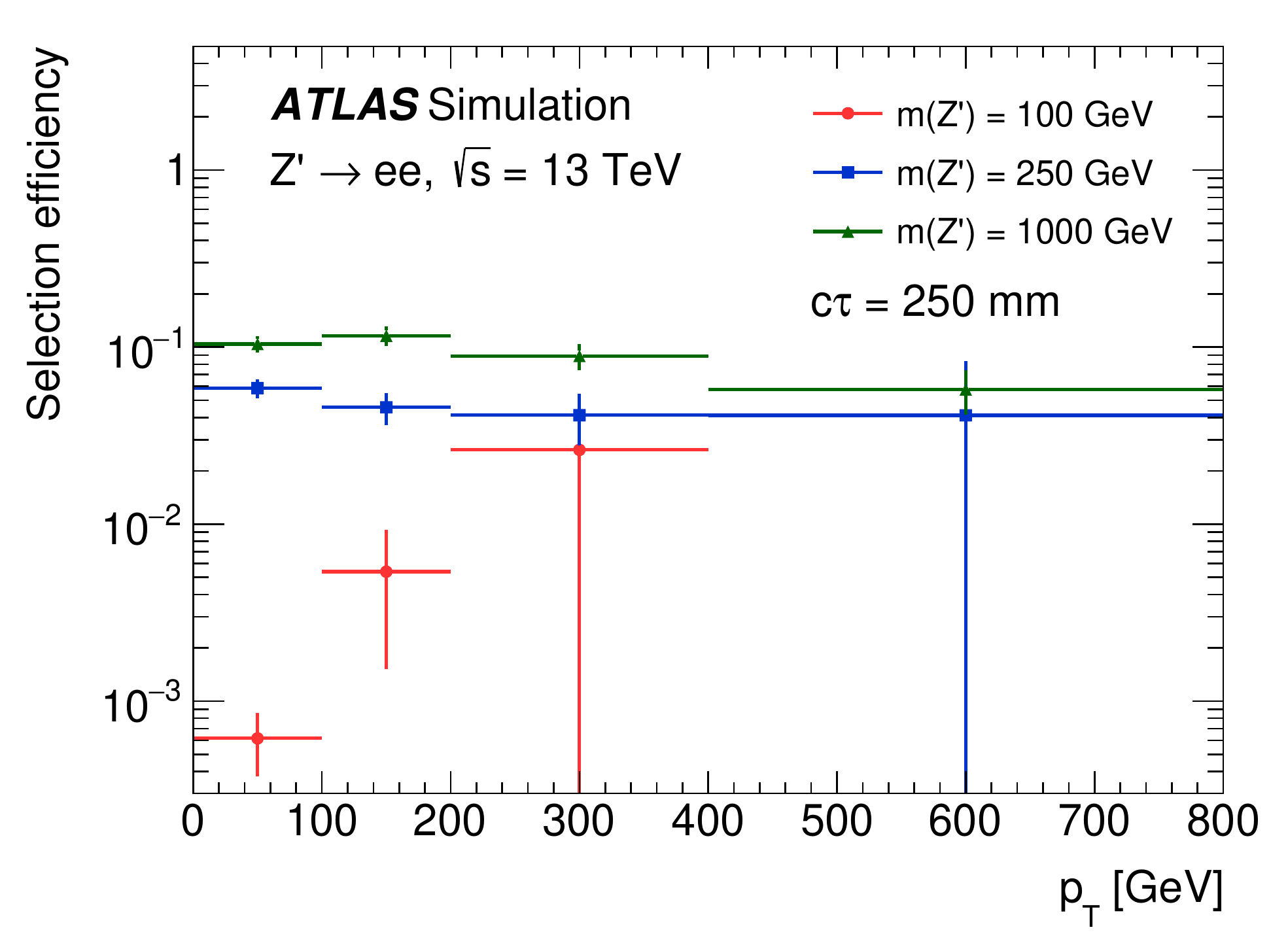}} \\
\subfloat{\includegraphics[width=0.5\textwidth]{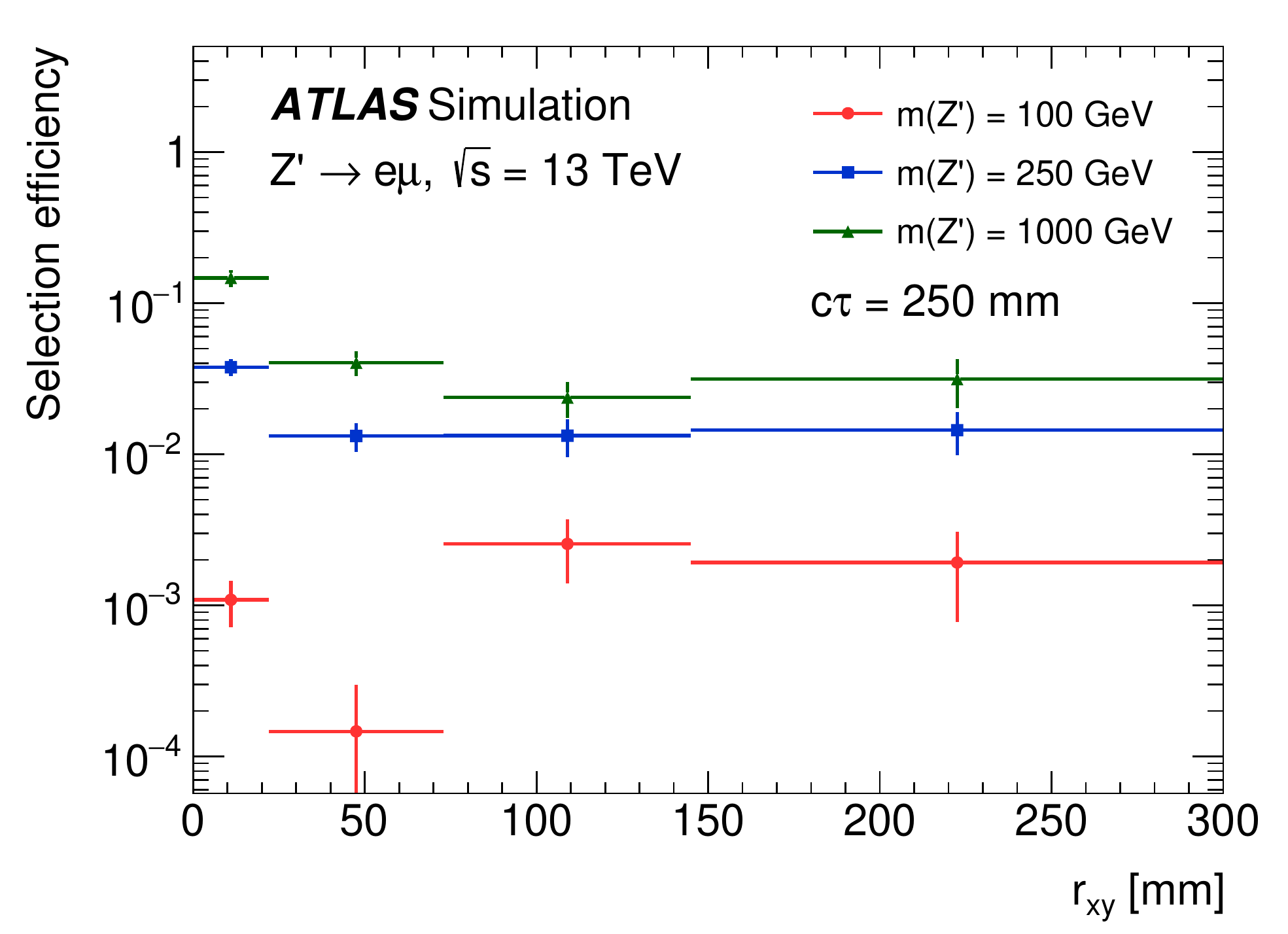}}
\subfloat{\includegraphics[width=0.5\textwidth]{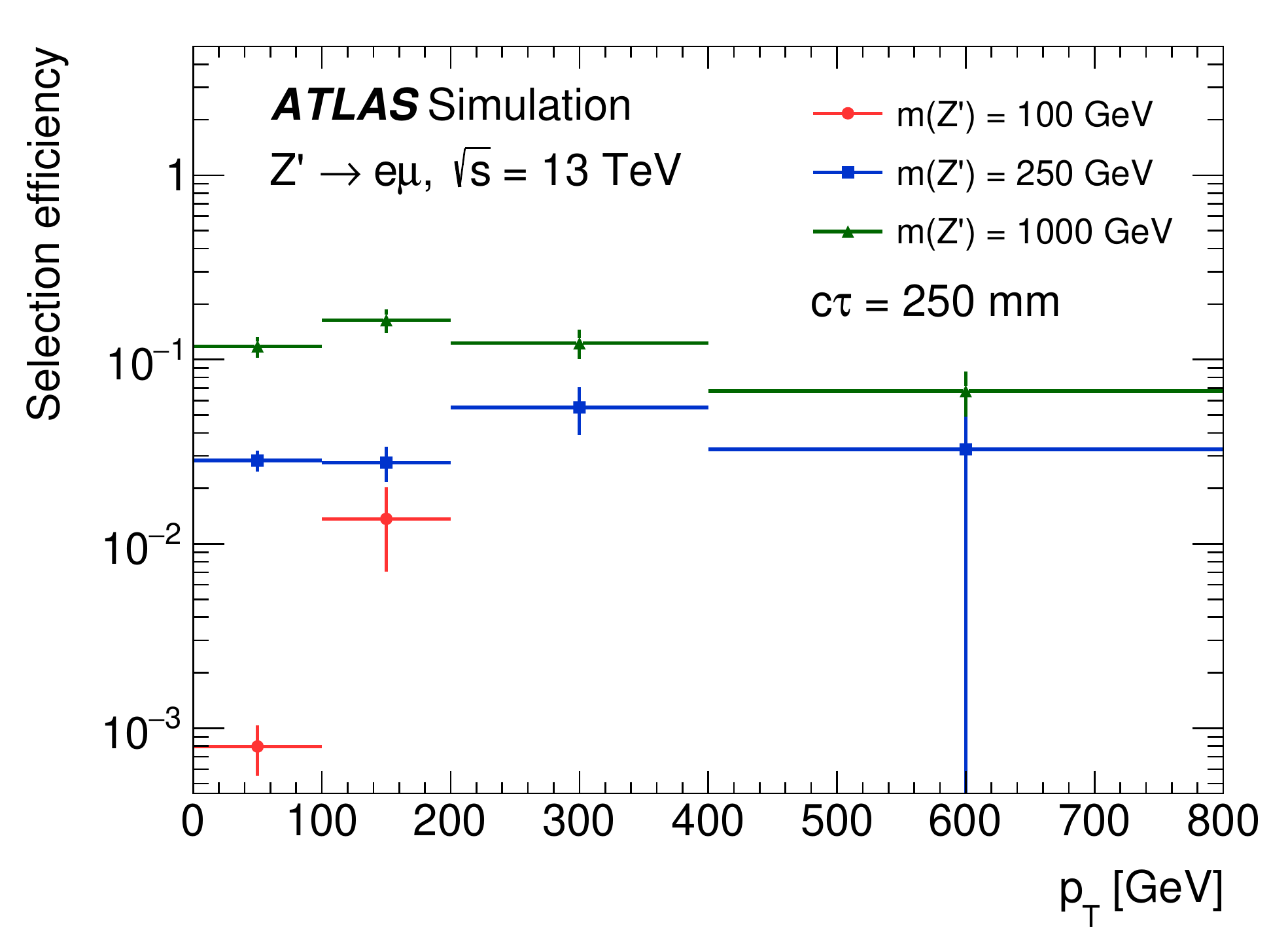}} \\
\subfloat{\includegraphics[width=0.5\textwidth]{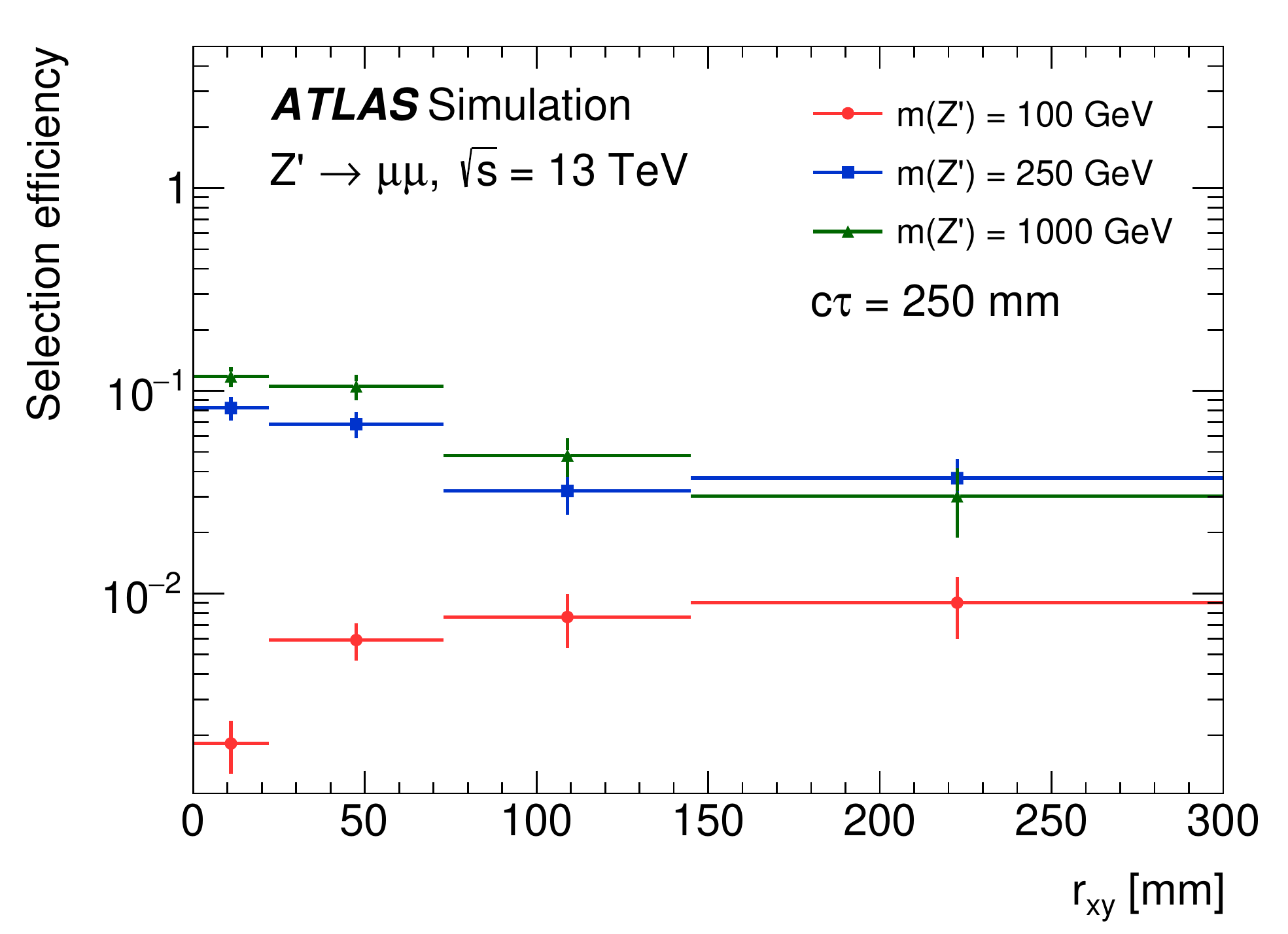}}
\subfloat{\includegraphics[width=0.5\textwidth]{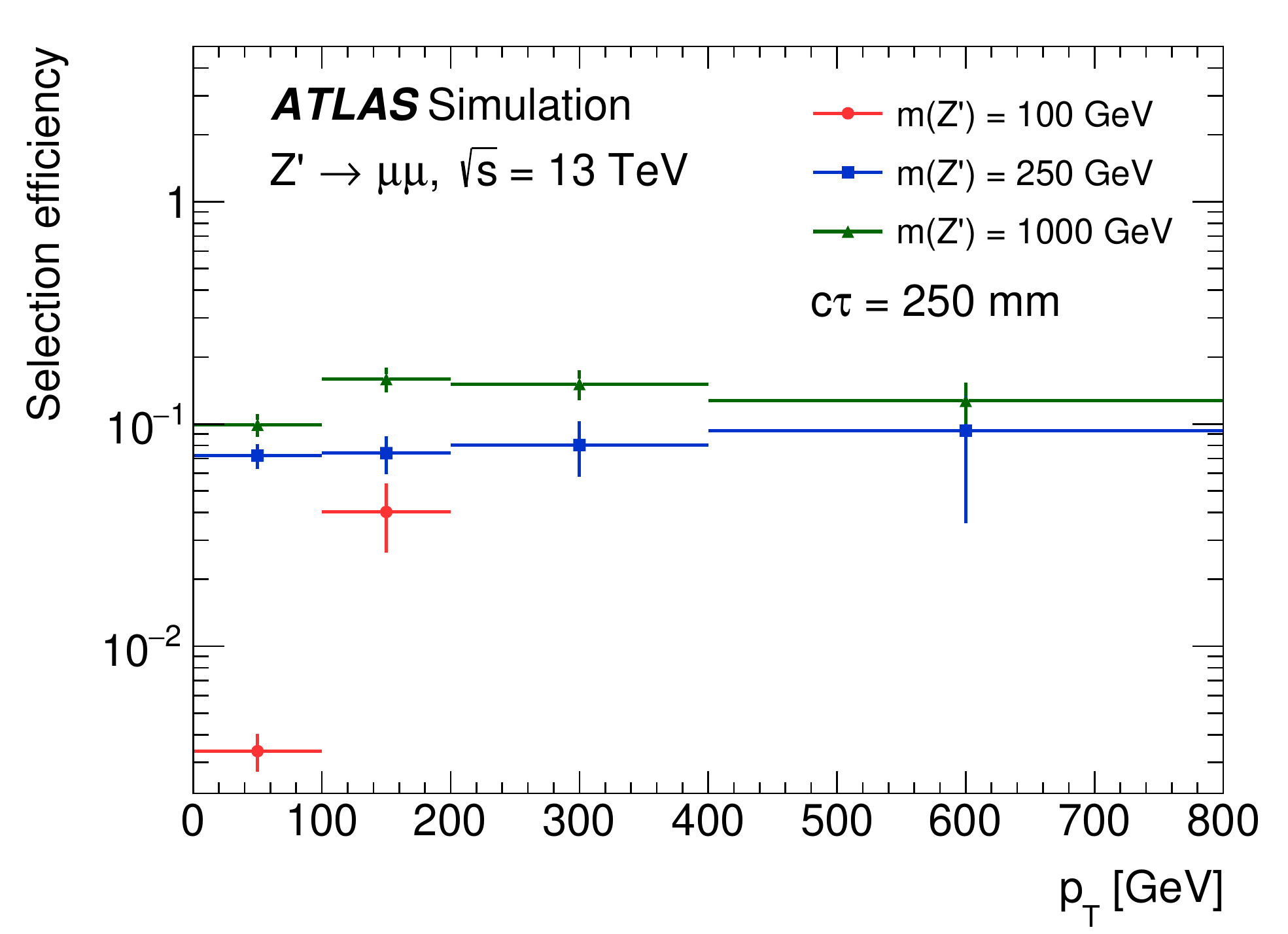}} \\
\caption{Dependence of the overall signal efficiency on (left) the $r_{xy}$ and (right) the \pt of the long-lived $Z'$ for three masses, $c\tau = 250$~mm and (top) $Z' \rightarrow ee$, (middle) $Z' \rightarrow e\mu$ and (bottom) $Z' \rightarrow \mu\mu$. The error bars indicate the total uncertainties.}
\label{fig:Zprime_SigEff}
\end{figure}

Overall signal efficiencies for the $Z'$ toy model are shown in Figure~\ref{fig:Zprime_SigEff}.
The modified reconstruction techniques used here significantly improve the signal sensitivity out to large decay radii, with decreasing sensitivity at larger decay radius.
The signal efficiency increases with the heavier LLP because the leptons are produced with higher $\pt$, resulting in higher trigger efficiencies.
This is particularly evident for the $Z'$ with mass of 100~GeV in which the efficiency is significantly lower.
 
Overall, the LRT and DV techniques used in this study preserve detection efficiencies at large decay radii, showing sensitivity for LLPs with decay lengths of the order of the ID dimensions.
The HepData entry~\cite{HEPDATA} associated with this publication has additional efficiency maps and material to reinterpret the results of this search in other models.
 
\begin{figure}[tbp]
\centering
\subfloat{\includegraphics[width=0.5\textwidth]{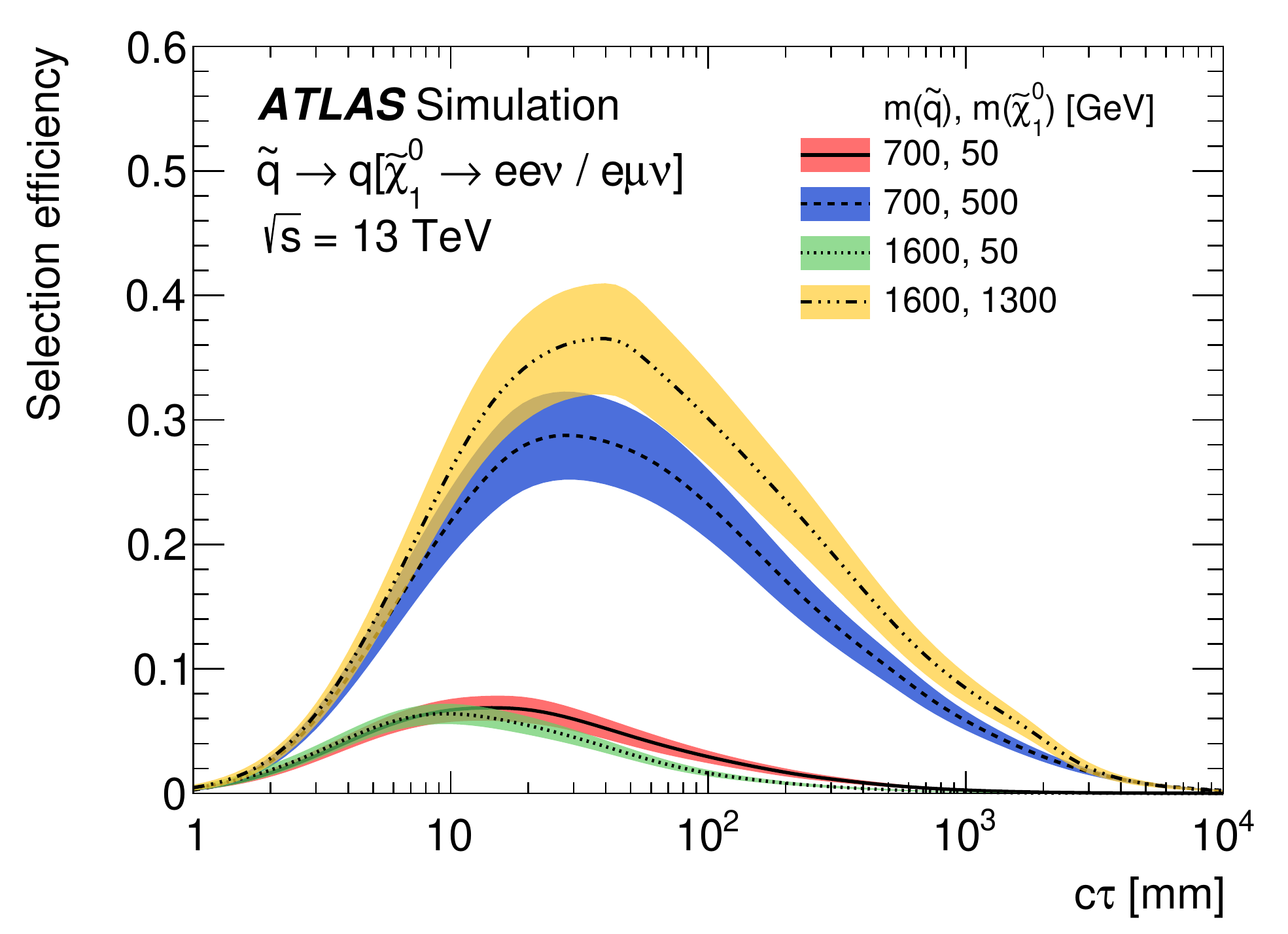}}
\subfloat{\includegraphics[width=0.5\textwidth]{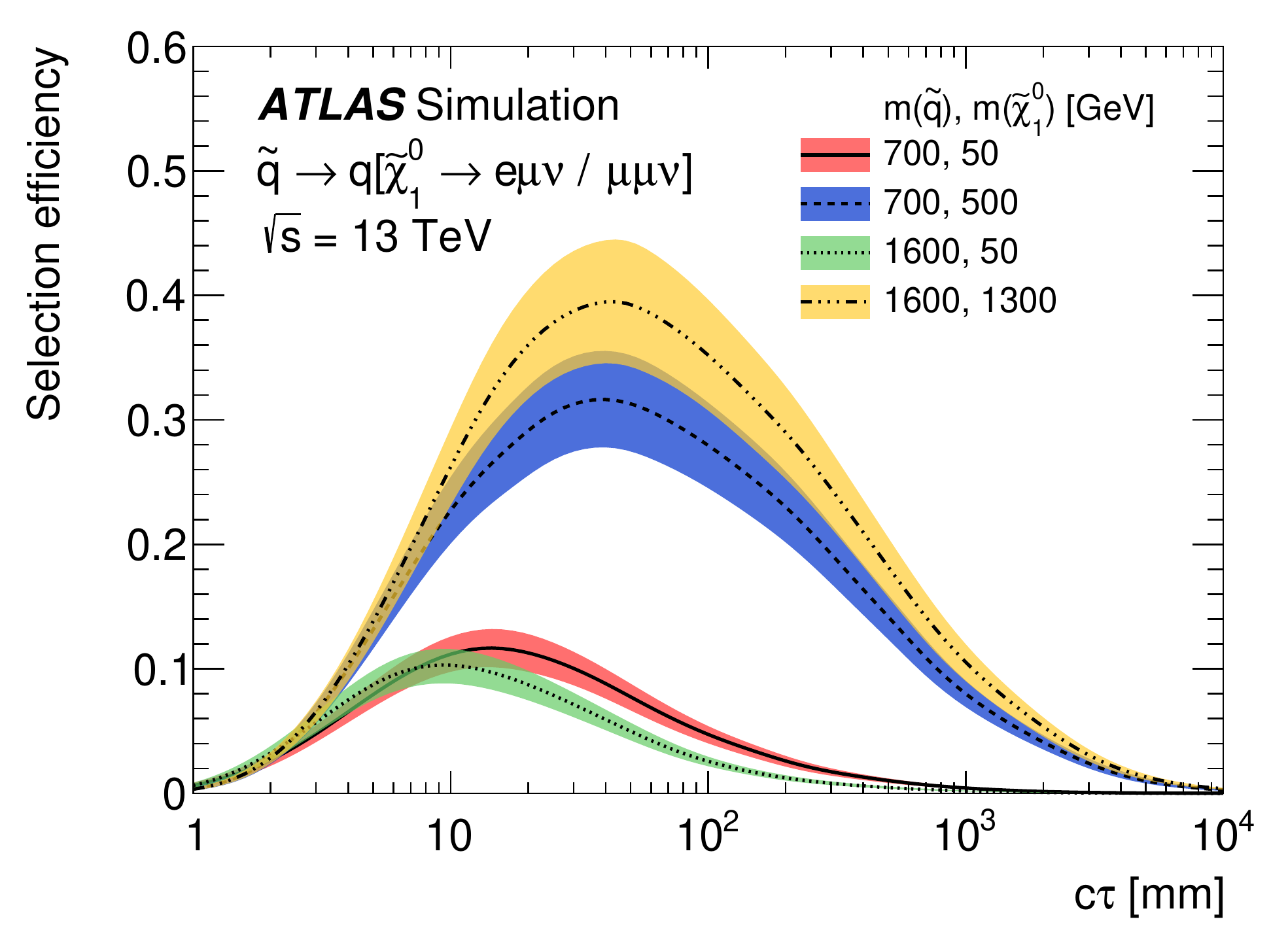}}
\caption{Overall signal efficiency as a function of the mean proper lifetime of the \ninoone, in units of $c\tau$, for (left) the $\lambda_{121}$ and (right) the $\lambda_{122}$ scenarios with various combinations of squark and \ninoone masses. The shaded bands indicate the total uncertainties.}
\label{fig:RPV_SigEvtEff}
\end{figure}

For the interpretations of the RPV SUSY model, two independent scenarios are considered, each corresponding to an LSP decay that is mediated by a single dominant RPV coupling, $\lambda_{121}$ or $\lambda_{122}$.
Figure~\ref{fig:RPV_SigEvtEff} shows the overall signal efficiencies for each scenario, which can be as high as $40\%$ for a 1.3~TeV neutralino and $12\%$ for a 50~GeV neutralino.
The selection criteria that have the largest impact on the signal efficiencies are the reconstruction efficiency of displaced vertices, the trigger and preselection requirements, and, in the case of $ee$ and $e\mu$ vertices, the material veto.
 
Upper limits at the 95\% confidence level (CL) are calculated with the \textit{HistFitter}~\cite{Baak:2014wma} package by testing various signal hypotheses using the $\text{CL}_{\text{s}}$ prescription~\cite{Read:2002hq} with a profile likelihood ratio as the test statistic, whose probability density function is determined with pseudo experiments.
 
The model-independent upper limit on the visible cross-section, which is defined as the production cross-section times the overall signal efficiency, $\langle \epsilon \sigma \rangle_{\text{obs}}^{\text{95}}$ is $0.09$~fb and on the visible number of signal events $S_{\text{obs}}^{\text{95}}$ is $3.0$~events, consistent with the expectation of $3.0^{+0.6}_{-0.0}$~events.
 
\begin{figure}[tbp]
\centering
\subfloat{\includegraphics[width = 0.49 \textwidth]{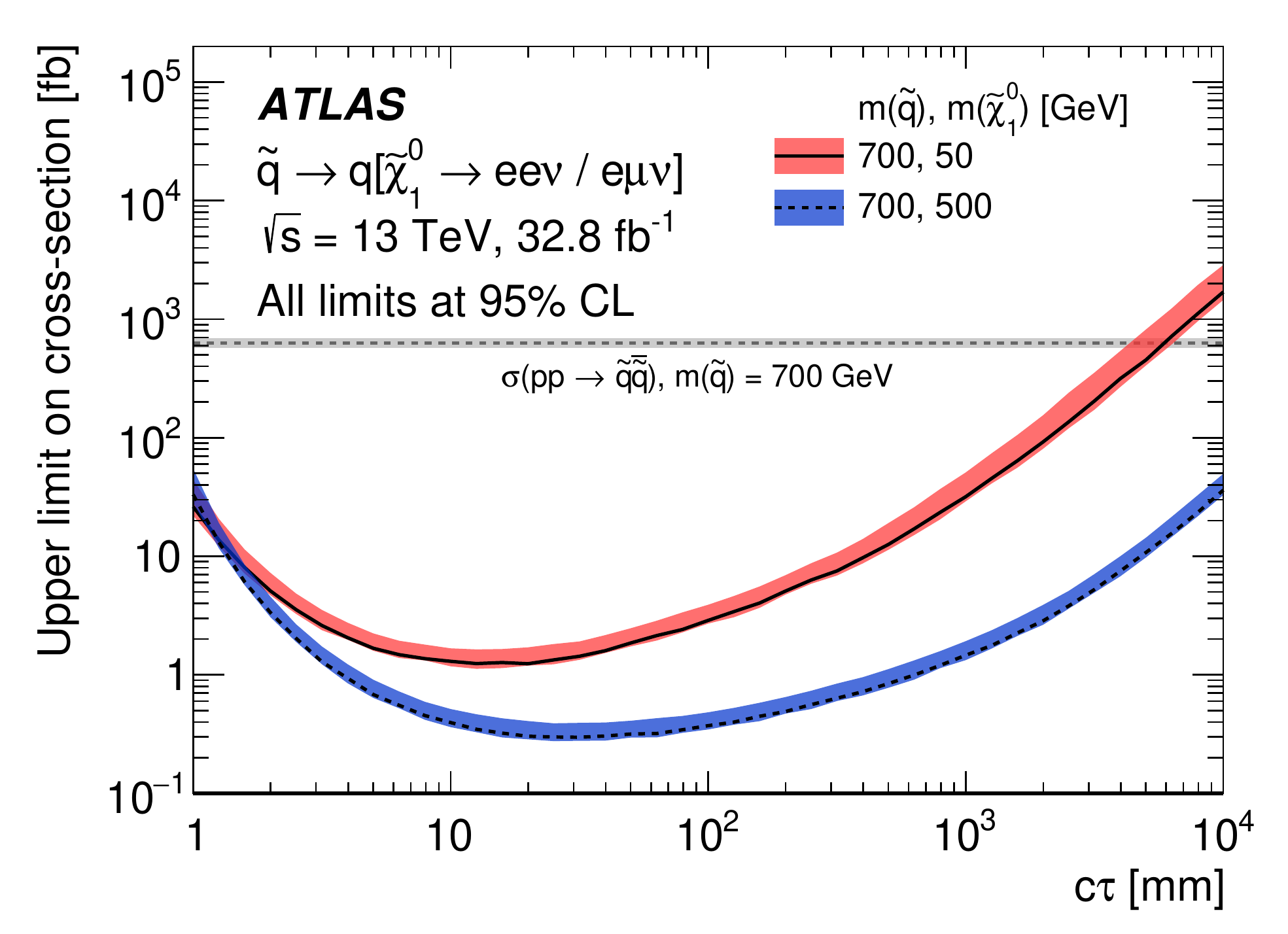}}
\subfloat{\includegraphics[width = 0.49 \textwidth]{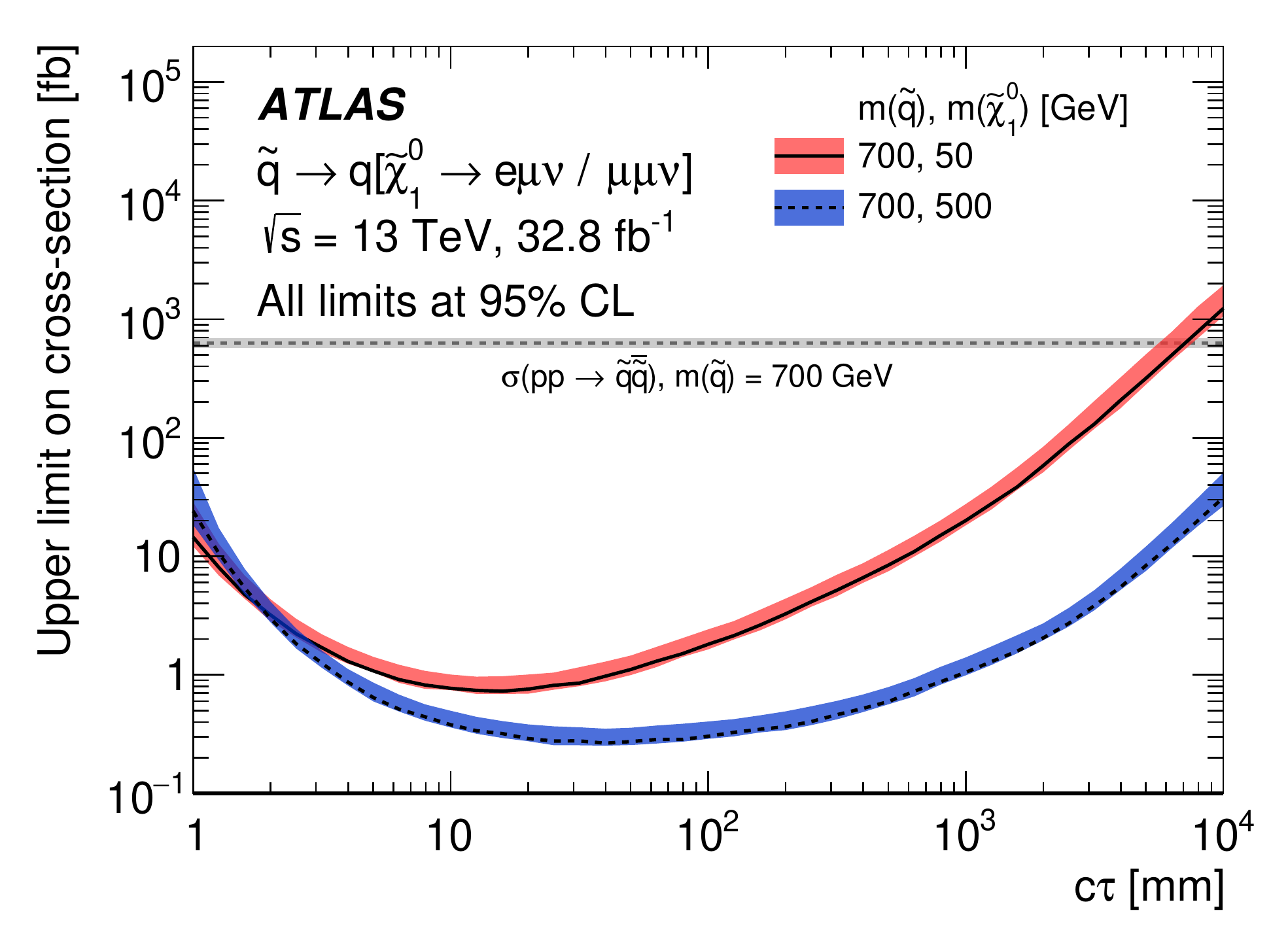}}
\caption{Observed upper limits at 95\% CL on the squark--antisquark production cross-section as a function of the mean proper lifetime of the \ninoone, in units of $c\tau$, for (left) the $\lambda_{121}$ and (right) the $\lambda_{122}$ scenarios and a 700~GeV squark. The horizontal lines indicate the theoretical (approx. NNLO+NNLL) cross-sections with the uncertainties shown as shaded bands. The shaded bands around the observed limits indicate the $\pm 1\sigma$ variations in the expected limit.}
\label{fig:RPV_CSlimits_700}
\end{figure}

\begin{figure}[tbp]
\centering
\subfloat{\includegraphics[width = 0.49 \textwidth]{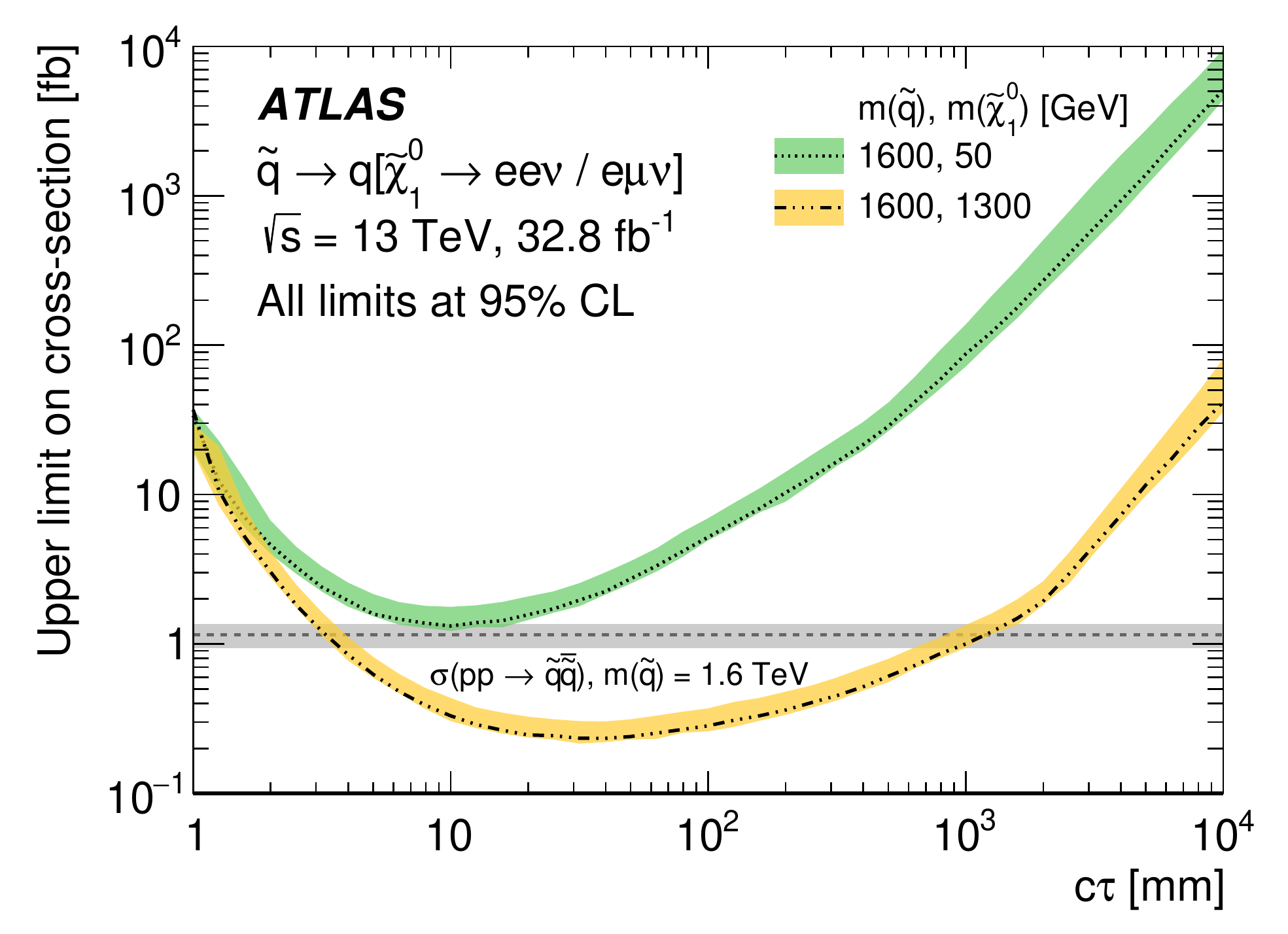}}
\subfloat{\includegraphics[width = 0.49 \textwidth]{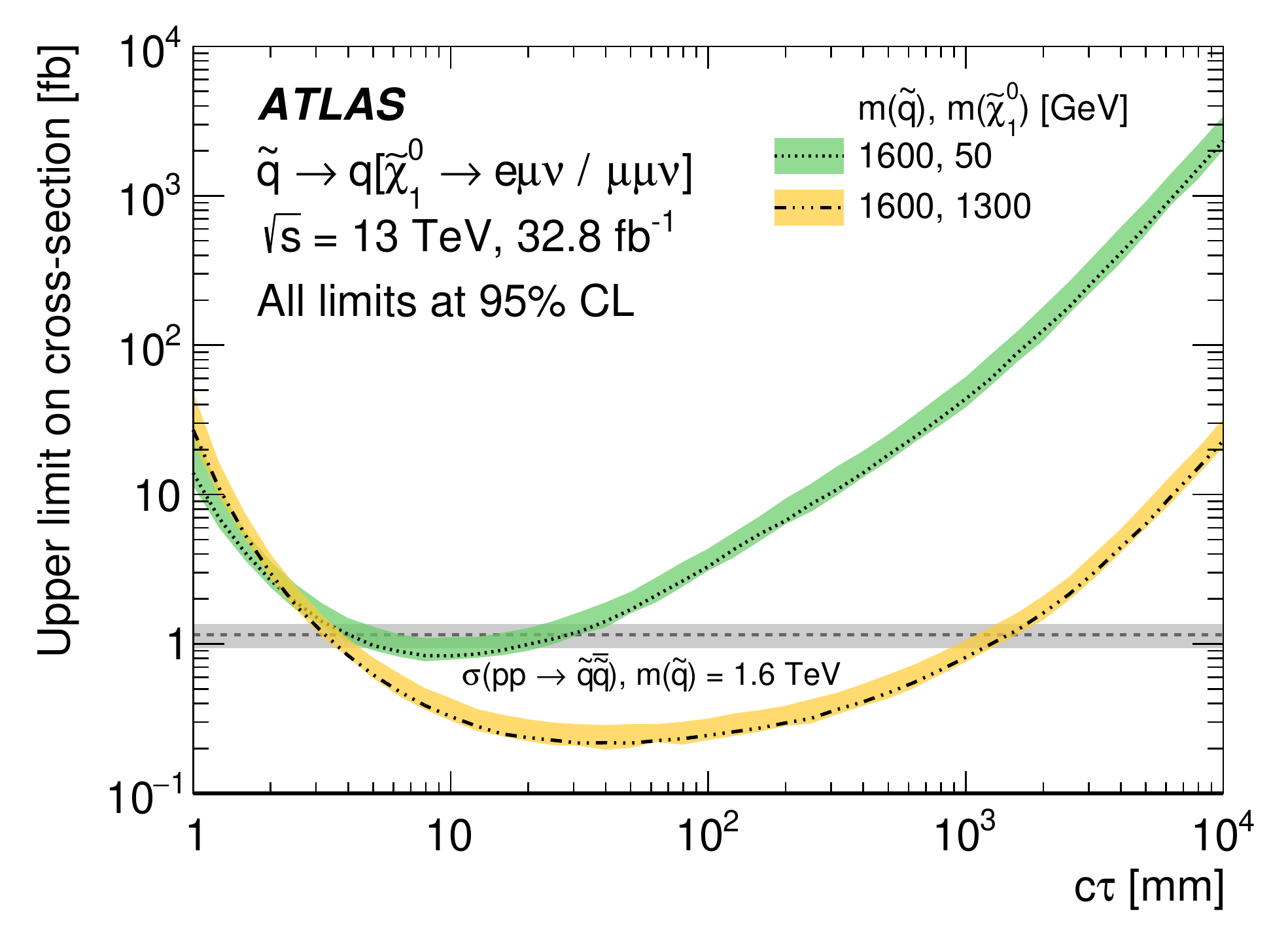}}
\caption{Observed upper limits at 95\% CL on the squark--antisquark production cross-section as a function of the mean proper lifetime of the \ninoone, in units of $c\tau$, for (left) the $\lambda_{121}$ and (right) the $\lambda_{122}$ scenarios and a 1.6~TeV squark. The horizontal lines indicate the theoretical (approx. NNLO+NNLL) cross-sections with the uncertainties shown as shaded bands. The shaded bands around the observed limits indicate the $\pm 1\sigma$ variations in the expected limit.}
\label{fig:RPV_CSlimits_1600}
\end{figure}

The upper limits on the production cross-section of the RPV SUSY model are shown in Figures~\ref{fig:RPV_CSlimits_700} and~\ref{fig:RPV_CSlimits_1600} for a 700~GeV squark and a 1.6~TeV squark, respectively.
The observed limits are consistent with the expected limits within the uncertainties.
Mean proper neutralino lifetimes which correspond to $c\tau$ between 1~mm and 6~m are excluded for a 700~GeV squark.
If the squark mass is much higher, 1.6~TeV, the excluded $c\tau$ region is 3~mm to 1~m for a 1300~GeV neutralino, while the $c\tau$ of a 50~GeV neutralino can be constrained only in the $\lambda_{122}$ scenario, where $c\tau$ values between 4 and 30~mm are excluded.
 
% End of text imported from the .//results.tex input file
 
\FloatBarrier
 
\section{Conclusion}
\label{sec:conclusion}
 
% The next lines are included from the .//conclusions.tex input file
A search for a long-lived particle with a mass of more than 12~GeV that decays into an oppositely charged $ee$, $\mu\mu$ or $e\mu$ pair within the ATLAS inner detector has been performed on 32.8~$\mathrm{fb^{-1}}$ of $pp$ collision data at $\sqrt{s}=13$~TeV collected in 2016.
No events are observed in the signal region, consistent with the expected background of $0.27 \pm 0.17$~events.
The detection efficiencies for generic resonances with lifetimes ($c\tau$) of 100--1000~mm decaying into a dilepton pair with masses between 0.1--1.0~TeV are presented as a function of $\pt$ and decay radius of the resonances to allow the extraction of upper limits on the cross sections for theoretical models.
Exclusion limits are derived for a supersymmetric signal model in which long-lived neutralinos (\ninoone) are produced through squark--antisquark production and decay into two charged leptons and one neutrino via R-parity violating couplings.
If the eight left- and right-handed squarks of the first two generations have a common mass of 700~GeV, mean proper lifetimes for long-lived neutralinos which correspond to $c\tau$ values between 1~mm and 6~m are excluded.
If the squark mass is 1.6~TeV, $c\tau$ values between 3~mm and 1~m are excluded for a 1.3~TeV neutralino, while for a 50~GeV neutralino the limits are significantly weaker.
% End of text imported from the .//conclusions.tex input file

\section*{Acknowledgements}
 
% The next lines are included from the .//acknowledgements/Acknowledgements.tex input file
 
We thank CERN for the very successful operation of the LHC, as well as the
support staff from our institutions without whom ATLAS could not be
operated efficiently.
 
We acknowledge the support of ANPCyT, Argentina; YerPhI, Armenia; ARC, Australia; BMWFW and FWF, Austria; ANAS, Azerbaijan; SSTC, Belarus; CNPq and FAPESP, Brazil; NSERC, NRC and CFI, Canada; CERN; CONICYT, Chile; CAS, MOST and NSFC, China; COLCIENCIAS, Colombia; MSMT CR, MPO CR and VSC CR, Czech Republic; DNRF and DNSRC, Denmark; IN2P3-CNRS, CEA-DRF/IRFU, France; SRNSFG, Georgia; BMBF, HGF, and MPG, Germany; GSRT, Greece; RGC, Hong Kong SAR, China; ISF and Benoziyo Center, Israel; INFN, Italy; MEXT and JSPS, Japan; CNRST, Morocco; NWO, Netherlands; RCN, Norway; MNiSW and NCN, Poland; FCT, Portugal; MNE/IFA, Romania; MES of Russia and NRC KI, Russian Federation; JINR; MESTD, Serbia; MSSR, Slovakia; ARRS and MIZ\v{S}, Slovenia; DST/NRF, South Africa; MINECO, Spain; SRC and Wallenberg Foundation, Sweden; SERI, SNSF and Cantons of Bern and Geneva, Switzerland; MOST, Taiwan; TAEK, Turkey; STFC, United Kingdom; DOE and NSF, United States of America. In addition, individual groups and members have received support from BCKDF, CANARIE, CRC and Compute Canada, Canada; COST, ERC, ERDF, Horizon 2020, and Marie Sk{\l}odowska-Curie Actions, European Union; Investissements d' Avenir Labex and Idex, ANR, France; DFG and AvH Foundation, Germany; Herakleitos, Thales and Aristeia programmes co-financed by EU-ESF and the Greek NSRF, Greece; BSF-NSF and GIF, Israel; CERCA Programme Generalitat de Catalunya, Spain; The Royal Society and Leverhulme Trust, United Kingdom.
 
The crucial computing support from all WLCG partners is acknowledged gratefully, in particular from CERN, the ATLAS Tier-1 facilities at TRIUMF (Canada), NDGF (Denmark, Norway, Sweden), CC-IN2P3 (France), KIT/GridKA (Germany), INFN-CNAF (Italy), NL-T1 (Netherlands), PIC (Spain), ASGC (Taiwan), RAL (UK) and BNL (USA), the Tier-2 facilities worldwide and large non-WLCG resource providers. Major contributors of computing resources are listed in Ref.~\cite{ATL-GEN-PUB-2016-002}.
 
% End of text imported from the .//acknowledgements/Acknowledgements.tex input file

\FloatBarrier
 
\printbibliography

\newpage
\input{atlas_authlist} 
\end{document}

%% file: atlas_authlist.tex
% ATLAS Collaboration author list
% Reference date of SUSY-2017-04 is 2019-04-08
% Author list last updated on date 20-FEB-20
% Data extracted on 20-Feb-2020 for paper reference SUSY-2017-04
% at 10:53pm
 
\begin{flushleft}
{\Large The ATLAS Collaboration}

\bigskip

G.~Aad$^\textrm{\scriptsize 102}$,    
B.~Abbott$^\textrm{\scriptsize 129}$,    
D.C.~Abbott$^\textrm{\scriptsize 103}$,    
A.~Abed~Abud$^\textrm{\scriptsize 71a,71b}$,    
K.~Abeling$^\textrm{\scriptsize 53}$,    
D.K.~Abhayasinghe$^\textrm{\scriptsize 94}$,    
S.H.~Abidi$^\textrm{\scriptsize 167}$,    
O.S.~AbouZeid$^\textrm{\scriptsize 40}$,    
N.L.~Abraham$^\textrm{\scriptsize 156}$,    
H.~Abramowicz$^\textrm{\scriptsize 161}$
% ...
\bigskip
\\

% ...
$^{40}$Niels Bohr Institute, University of Copenhagen, Copenhagen; Denmark.\\
% ...
$^{53}$II. Physikalisches Institut, Georg-August-Universit\"{a}t G\"ottingen, G\"ottingen; Germany.\\
% ...
$^{71}$$^{(a)}$INFN Sezione di Pavia;$^{(b)}$Dipartimento di Fisica, Universit\`a di Pavia, Pavia; Italy.\\
% ...
$^{94}$Department of Physics, Royal Holloway University of London, Egham; United Kingdom.\\
% ...
$^{102}$CPPM, Aix-Marseille Universit\'e, CNRS/IN2P3, Marseille; France.\\
$^{103}$Department of Physics, University of Massachusetts, Amherst MA; United States of America.\\
% ...
$^{129}$Homer L. Dodge Department of Physics and Astronomy, University of Oklahoma, Norman OK; United States of America.\\
% ...
$^{156}$Department of Physics and Astronomy, University of Sussex, Brighton; United Kingdom.\\
% ...
$^{161}$Raymond and Beverly Sackler School of Physics and Astronomy, Tel Aviv University, Tel Aviv; Israel.\\
% ...
$^{167}$Department of Physics, University of Toronto, Toronto ON; Canada.\\
% ...
$^{a}$ Also at Borough of Manhattan Community College, City University of New York, New York NY; United States of America.\\
$^{b}$ Also at CERN, Geneva; Switzerland.\\
% ...

\end{flushleft}

% Created with Glance <Atlas.Glance@cern.ch>